\documentclass[%
 reprint,
 amsmath,amssymb,
 aps,
]{revtex4-2}

\usepackage{amsmath,amsfonts,amssymb,graphicx,braket,mathtools,array}
\usepackage{float}
\usepackage{multirow}
\usepackage[]{units}
\usepackage{hyperref}
\usepackage{MnSymbol}
\hypersetup{
    colorlinks=true,
    linkcolor=blue,
    filecolor=blue,      
    urlcolor=blue,
    citecolor=blue
}

\usepackage{natbib}
\usepackage{graphicx}
\usepackage[font=small,
   justification=Justified,
   format=plain]{caption}
\usepackage{dcolumn}
\usepackage{bm}
\usepackage[dvipsnames]{xcolor}
\usepackage[caption=false]{subfig}
\usepackage{mathrsfs}
\usepackage{amsmath}
\DeclareMathOperator{\Span}{span}

\begin{document}
\title{Fluctuating parametric drive of coupled classical oscillators can simulate dissipative qubits}
\author{Lorenzo Bernazzani}
\email{lorenzo.bernazzani@uni-konstanz.de}
\affiliation{Department of Physics, University of Konstanz, D-78457 Konstanz, Germany.}
\author{Guido Burkard}
\email{guido.burkard@uni-konstanz.de}
\affiliation{Department of Physics, University of Konstanz, D-78457 Konstanz, Germany.}


\begin{abstract}
We investigate a system composed of two coupled oscillators subject to stochastic fluctuations in its internal parameters. In particular, we answer the question whether the well-known classical analogy of the quantum dynamics of two-level systems (TLS), i.e. qubits, provided by two coupled oscillators can be extended to simulate the dynamics of dissipative quantum systems. In the context of nanomechanics, the analogy in the dissipation free case has already been tested in multiple experimental setups, e.g., doubly clamped or cantilever string resonators and optically levitated particles. A well-known result of this classical analogy is that the relaxation and decoherence times of the analog quantum system must be equal, i.e. $T_1=T_2$, in contrast to the general case of quantum TLS.  We show that this fundamentally quantum feature, i.e. $T_1\neq T_2$, can be implemented as well in the aforementioned classical systems by adding stochastic fluctuations in their internal parameters. Moreover, we show that these stochastic contributions can be engineered in the control apparatus of those systems, discussing, in particular, the application of this theory to levitated nanoparticles and to nanostring resonators.
\end{abstract}

\maketitle

\section{Introduction}

Enquiries on quantum-classical analogies have attracted much interest  \cite{Grover2002,Shore2009,Briggs2012,Briggs2013,Rodriguez2016}, and it is well established that the dynamics of a quantum N-level system described by Schr\"odinger’s equation can be simulated by classical systems, e.g. coupled classical harmonic oscillators \cite{Spreeuw90,Novotny2010,Skinner2013,Novotny2014,Novotny2016,Frimmer2017,Nori2018,EichlerPRL2022,Kiselev2022}. To date, several papers have outlined that this similarity between the two dynamics leads to the observation of purely quantum mechanical effects in classical systems, e.g. Rabi oscillations \cite{Spreeuw90,Weig2013,Novotny2017}, Landau-Zener transitions \cite{Weig2012}, and also St\"uckelberg interferometry \cite{Seitner2016,Fu2016,Seitner2017,Nori2018}. Quite surprisingly, some of these effects have been reported also in macroscopic systems \cite{Kiselev2022}, pushing this analogy even out to the macroscopic world.
As a consequence, classical coupled oscillators constitute the simplest platform to study how a classical system can mimic quantum dynamics.

Mechanics was indeed the first playground of physics. Nowadays, it is, maybe surprisingly, still investigated in the modern quest for the nanoscale miniaturization of physical devices \cite{RoukesBK,Eichler2022}. An astounding degree of control has been achieved in mechanical systems over the past years \cite{VDZant2012,Quidant2016,NovotnyLevi2021,DykmanRMP2022}. This has fostered the hope \cite{AspelmeyerPT,Stickler2020} to reach mesoscopic quantum superposition of massive objects and to study quantum effects of gravity in the lab \cite{AspelmeyerRMP2014,Ulbricht2017,NovotnyLevi2021}, with the aim of uncovering some of the elusive aspects of the quantum-classical border \cite{Arndt2014,Stickler2020}.

This work is meant to give some new insights about the quantum-classical analogy, drawing from the aforementioned mapping of quantum evolution by means of classical oscillators. In particular, we address the issue of the trivial form of the relaxation term \cite{Weig2013,Novotny2014,Novotny2017,Nori2018} in the Schr\"odinger equation for the simulated quantum TLS, i.e., the fact that all the components of the Bloch vector (BV) decay with the same characteristic time. It is well known instead, that in the quantum case there are two relaxation times, the longitudinal $T_1$ and the transverse $T_2$. These are linked to the relaxation of the populations and of the coherences of the TLS state. Furthermore, they are related by the equation $T^{-1}_2=(2T_1)^{-1}+T^{-1}_{\phi}$, which also defines the phase relaxation time $T_{\phi}$ \cite{Burkard2008}. Our aim is to show how noise can induce dissipation mechanisms with such quantum features, that were not grasped by the previous classical model. The addition of stochastic fluctuations and the solution of the associated Langevin dynamics \cite{Chandrasekhar43} adds quantumness to the system, meaning that it provides the phenomenology of this model with a pure phase relaxation time, dependent on the noise strength. 

The quantum phase is the defining concept of quantum mechanics, giving rise to quantum superpositions, interference phenomena, and many-body entanglement. The perturbation of microscopic quantum systems leads to the loss of phase coherence, and, on a macroscopic scale, the emergence of our classical reality. Decoherence is then a detrimental aspect for quantum information systems \cite{Burkard2008}. However, the interaction of quantum systems with the external degrees of freedom of the environment is unavoidable. The attempt to suppress the coupling between the system and environment has been complemented with the modelling of the effects of these interaction on the reduced system dynamics. While an accurate model can be achieved in simple systems, the computational complexity scales exponentially with the system size. To cope with this, simulation of dissipation is a valuable tool \cite{Daley2014}. The common approach consists in adding classical noise to the analog system. In this way the open-quantum-system, that one is interested in simulating, is mapped onto another system, more controllable, the dynamics of which is proven to be analog.  Our work also follows this direction by providing a very simple classical analog system, in which aspects of quantum dissipation can be simulated and even visualized \cite{Kiselev2022}.
The approach to analog simulation of quantum dissipation via classical noise has a precedent in the context of many-body quantum systems \cite{DelCampo2017}, and the relation between classical and quantum noise in open systems dynamics has been thoroughly investigated and described \cite{Burgarth2017,Franco2019,Jordan2022}.
The novelty of this paper lies in the fact that the system is completely classical. Therefore it constitutes a furthest simplification of the aforementioned results that, moreover, can be readily tested in the lab.

\section{The classical two-level atom}

\begin{figure}[t]
\begin{center} \includegraphics{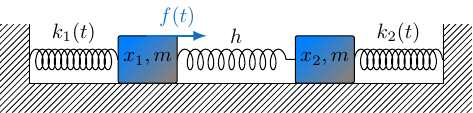}
\end{center}
\caption{Schematics of the classical-mechanical model under consideration. Two masses $m$ are connected through a spring with the time-independent spring constant $h$, and each of them is connected to the neighbouring wall by springs with time-dependent spring constants $k_1(t)$ and $k_2(t)$, with $k_2(t)-k_1(t)=\Delta k(t)\,$. We also include friction by means of a damping coefficient $\gamma$ equal for both masses. Moreover, we include an external driving force $f(t)$. This external driving has the purpose of initially feeding energy into the system that is then driven parametrically \cite{Rugar91,DykmanRMP2022}, via the modulation of the spring constants.}
\label{Fig1}
\end{figure}

Let us start from a system of coupled classical mechanical oscillators consisting of two bodies with equal mass $m$ connected with springs to each other and to the adjacent fixed walls, as depicted in Fig.~\ref{Fig1}. This system is described by the following linear coupled classical equations of motion
for the positions $x_1$ and $x_2$ of the two masses,
\begin{align}
m\ddot{x}_1 + m\gamma\dot{x}_1 + \biggl[k-\frac{\Delta k(t)}{2}\biggr]x_1+h\bigl(x_1-x_2\bigr)&=f(t),\\
m\ddot{x}_2 + m\gamma\dot{x}_2 + \biggl[k+\frac{\Delta k(t)}{2}\biggr]x_2+h\bigl(x_2-x_1\bigr)&=0\,,
\label{EQMot}
\end{align}
where we assumed that the masses $m$ and the damping coefficient $\gamma$ of the two oscillators are equal for convenience. The time dependent spring constants $k_{1,2}(t)=k\mp \Delta k(t)/2$ here account for the parametric driving mechanism that we are going to discuss extensively later. The (linear) coupling between the two oscillators is given by $h$. The inhomogeneous term on the right hand side of the first equation describes the effect of an external time-dependent deterministic or fluctuating force. This term has the purpose of injecting energy into the system by displacing the oscillator $x_1$ from its equilibrium position. We discuss later what happens when we let the system evolve once it has been initialized in a certain state. Therefore, in the following we consider the time-evolution of the system after the initial time $t=0$ at which the inhomogeneous force stops.
In the rest of the paper, we set then the inhomogeneus term $f(t)=0$. Nonetheless, since we have introduced friction through the $\gamma$ coefficient, it would generally be the case that a fluctuating force of thermal origin will also appear in $\delta f(t)$. For simplicity, we delegate the treatment of the more general case $\delta f(t)\neq 0$ to Appendix~\ref{sec:non-hom}. In any case, the suppression of this thermal noise term can be accomplished with little effort in real systems by lowering the temperature of the thermal bath or by using feedback mechanisms \cite{Poggio2007,VDZant2012,Quidant2016}.

Following Ref.~\onlinecite{Novotny2014}, let us now divide Eq.~\eqref{EQMot} by $m$ and introduce the quantities $\omega_0^2\equiv\frac{k+h}{m}\,,\,\Omega^2_c\equiv\frac{h}{m}\,,\,\Omega^2_d(t)\equiv\frac{\Delta k(t)}{2m}\,$.
After this relabeling we can rewrite the Eq.~\eqref{EQMot} in the following matrix form,
\begin{equation}
\Bigl(\frac{d^2}{dt^2}+\gamma\frac{d}{dt}
+\omega_0^2\Bigr)\begin{bmatrix}
x_1 \\
x_2
\end{bmatrix}+
\begin{bmatrix}
-\Omega^2_d & -\Omega^2_c \\
-\Omega^2_c & \Omega^2_d
\end{bmatrix}\begin{bmatrix}
x_1 \\
x_2
\end{bmatrix}=\begin{bmatrix}
0 \\
0
\end{bmatrix}.
\end{equation}
Assuming that the oscillators are not over-damped, we are now going to make the ansatz $x_{1,2}(t)=\mathrm{Re}{\bigl[\psi_{1,2}(t)\,e^{i\omega_0t}\bigr]}$. Basically this consists in factorising the oscillation into an oscillating component at the carrier frequency $\omega_0$ times an amplitude modulation $\psi_{1,2}(t)$. If we make the further assumption that the envelope $\psi_{1,2}$ is a slowly varying function of time we can drop the second derivative in our equations of motion. This has been called the slowly varying envelope approximation (SVEA) \cite{Novotny2014}. Furthermore, we assumed $\gamma\ll \omega_0$ \cite{Novotny2014,Nori2018}, which is typically the case in state of the art micro/nanomechanics since $\gamma=\omega_0/Q$, where $Q$ is the quality factor, which usually varies between $10^3$ and $10^6$ for most of the systems of interest \cite{VDZant2012}. These modifications lead to a matrix equation for the complex valued amplitudes alone, written in a vectorial form, i.e., utilizing the vector $\psi=\bigl[\,\psi_1\,,\,\psi_2\,\bigr]^T$,
\begin{equation}
i\dot{\psi}=\mathrm{H}(t)\psi-i\frac{\gamma}{2}\psi=\,\frac{1}{2}\bigl[\Delta\sigma_x+\varepsilon(t)\sigma_z\bigr]\psi-i\frac{\gamma}{2}\psi\,,
\label{EQTLS}
\end{equation}
where $\Delta\equiv\Omega^2_c/\omega_0$ and $\varepsilon\equiv\Omega^2_d/\omega_0$.
In the absence of friction $\gamma=0$, the two-component-amplitude equation \eqref{EQTLS}  is an analogy of the Schr\"odinger equation for a driven TLS (taken with $\hbar=1$) \cite{SchevchenkoBK}.

It is customary now to introduce a harmonic form for the driving term and we will specifically use $
\varepsilon(t)=\varepsilon_0+D\cos{\omega t}\,$.
It has  already been shown that this parametric driving induces a coherent dynamics that is quite faithfully resembling a quantum TLS. In fact, we could introduce either density matrix or Bloch vector (BV) equivalent representations \cite{Feynman57,EberlyBK,AbragamBK},
with the definitions,
\begin{equation}
\rho=\begin{bmatrix}
|\psi_1|^2 & \psi^{\ast}_1\psi_2 \\
\psi^{\ast}_2\psi_1 & |\psi_2|^2
\end{bmatrix}\,,\quad\boldsymbol{\mathrm{r}}=\begin{bmatrix}
\psi^{\ast}_1\psi_2 + \psi^{\ast}_2\psi_1\\
i\bigl(\psi^{\ast}_2\psi_1-\psi^{\ast}_1\psi_2\bigr)\\
|\psi_2|^2-|\psi_1|^2.
\end{bmatrix}
\label{Bloch}
\end{equation}
It must be noted however that neither, in the general case, $\psi$ and $\rho$ have modulus or trace equal to $1$ nor the dynamical map of Eq.~\ref{EQTLS} is trace preserving. The modulus or trace of the state would be dependent on the energy fed into the system by the inhomogeneous driving $f(t)$, for $t<0$. From the point $t=0$ onward anyway, the only source of damping of energy (probability in the quantum analogy) will be the non-hermitian trivial part of the Hamiltonian. We are going to get rid of this part in the next section by a simple coordinate transformation. Notwithstanding, the Hamiltonian with $\gamma=0$ is conservative in the classical language and Hermitian in the quantum analogy, therefore that part is energy/trace conserving. Standing these \emph{caveat}, we call our $\psi$ and $\rho$ states nonetheless.

Since the Hamiltonian in Eq.~\eqref{EQTLS} can be written in the form $\mathrm{H}(t)=\boldsymbol{\mathrm{B}}(t)\cdot\boldsymbol{\sigma}$ this quantum classical analogy leads to the so-called classical Bloch equations (BE) \cite{Novotny2014}, i.e.,
\begin{equation}
\boldsymbol{\dot{\mathrm{r}}}=\boldsymbol{\mathrm{B}}(t)\times \boldsymbol{\mathrm{r}}-\gamma\boldsymbol{\mathrm{r}}\,.
\label{BlochEOM}
\end{equation}
Here $\boldsymbol{\mathrm{B}}(t)=\bigl[\,\Delta\,,\,0\,,\,\varepsilon(t)\,\bigr]^T$ is the magnetic field vector and its modulus $B$, is the angular velocity with which the Bloch vector precesses. In a magnetic resonance description this describes the Larmor precession of a spin with Bloch vector ${\bf r}$ around a magnetic field along $\boldsymbol{\mathrm{B}}$.

The resemblance of quantum dynamics shown by Eq.~\eqref{BlochEOM} is striking considered that they have been derived by purely classical means. However, there is one evident issue in these equations, which makes the classical simulation less faithful, i.e., that the relaxation terms are not equal to the general relaxation present, e.g., in dissipative spin dynamics. More precisely, we know from the theory of either light-matter interaction or magnetic resonance that the components of the BV relax with two characteristic times \cite{AbragamBK,EberlyBK,Burkard2008}: $T_2$ is the relaxation time of the $x$ and $y$ components of the BV, hence the ones containing the coherences, while $T_1$ is the relaxation time for the $z$ component, related to level populations in the quantum language. In the following, we will explain how this hallmark quantum feature can be implemented in the aforementioned analogy.

\section{Fluctuations in the drive}

We make a change of coordinates to go into the \emph{diabatic} basis, i.e. $
\psi'=
\begin{bmatrix}
\psi_+\\
\psi_-
\end{bmatrix}
\equiv
\exp\Bigl(\frac{i\theta}{2}\sigma_y\Bigr)\begin{bmatrix}
\psi_1\\
\psi_2
\end{bmatrix}.$
After this rotation by the angle $\theta=\arctan{[\Delta/\varepsilon_0]}$, the Schr\"odinger equation takes the form
\begin{equation}
i\dot{\psi}'=\frac{1}{2}\biggl[\Omega\sigma_z+\frac{\varepsilon_0}{\Omega}\varepsilon'(t)\sigma_z-\frac{\Delta}{\Omega}\varepsilon'(t)\sigma_x\biggr]\psi'-i\frac{\gamma}{2}\psi'\,,
\label{EQDrivQubit}
\end{equation}
where $\varepsilon'(t)=D\cos(\omega t)$ and $\Omega=\sqrt{\varepsilon_0^2+\Delta^2}$.
It is also useful to renormalize the state vector in order to get rid of the trivial decay with rate $\gamma/2$ and thus to ensure a normalized state vector, as costumarily required by the quantum analogy. Therefore, we rename $\psi_{\pm}\to \mathcal{N}\psi'_{\pm}\,e^{-\gamma t/2}$, where $\mathcal{N}[E(0)]$  depends on $E(0)=\int_{-\infty}^{0}f(t)\dot{x}_1(t)\,dt$, i.e., the energy fed into the system before our simulation has started.

Our aim is to add fluctuations to the dynamics of the coupled oscillators system. We accomplish this by adding a stochastic term to the driving. Along this route, we write the driving term $\varepsilon(t)$ with an additive Langevin force (which actually is a multiplicative noise term for the stochastics problem \cite{RiskenBK}, since it multiplies a random variable),
\begin{equation}
\varepsilon'(t)=D\cos{(\omega t)}+\Gamma_d(t)\,.
\end{equation}
Here, $\Gamma_d$ has the dimension of a frequency.
We consider for the noise term a stationary and Gaussian Ornstein-Uhlenbeck process \cite{Ornstein30,Chandrasekhar43,RiskenBK} such that:
\begin{equation*}
\langle\Gamma_d(t)\rangle=0\,,\qquad\langle\Gamma_d(0)\Gamma_d(t)\rangle=\,\frac{G}{\tau_c}\exp{\biggl(-\frac{|t|}{\tau_c}\biggr)}\,,
\end{equation*}
where $G$ is the noise strength.

Therefore, we have to solve the following coupled stochastic differential equations \cite{VanKampenBK},
\begin{align}
i\dot{\psi}&=\Bigl[\mathrm{H}_0+\mathrm{H}_1(t)\Bigr]\psi\label{EQStochastic}\\
&=\,\frac{1}{2}\biggl[\Omega\sigma_z+\frac{D\cos{(\omega t)}+\Gamma_d(t)}{\Omega}\bigl(\varepsilon_0\sigma_z-\Delta\sigma_x\bigr)\biggr]\psi\nonumber\,,
\end{align}
where $\mathrm{H}_0=\frac{1}{2}\Omega\sigma_z$ and $\mathrm{H}_1(t)=\mathrm{H}_d(t)+\mathrm{H}_s(t)$ contains the time-dependent parts of the Hamiltonian, both deterministic $\mathrm{H}_d(t)\propto\cos(\omega t)$ and stochastic $\mathrm{H}_s(t)\propto\Gamma_d(t)$.
In order to solve this system of stochastic equations we rely on the cumulant expansion method. This method was pioneered by Kubo \cite{Kubo63} and then refined by others \cite{Fox75,Fox76,VanKampen74I,Terwiel74,Fox86,Skinner87,Skinner90}. These tools have been applied to magnetic resonance in molecular samples \cite{Freed68}. Particularly useful will be also the extension to coloured noise with correlations between the longitudinal and transverse component \cite{Leporini92,Leporini95}.

The aforementioned stochastic formalism allows one to replace the full stochastic differential equation with a differential equation for the averages of the stochastic variables, or of their higher moments. Since we are interested in the motion of the BV we will need the second moments of $\psi_{\pm}$. This leads us straightforwardly to the stochastic analog of the BE.
Since we need to get to an equation of motion for the avarage of the BV, this will contain the second moments of the stochastic variables $\psi_\pm$. The effective density matrix, with obvious changes from Eq.~\ref{Bloch} is now $\rho(t)=\sum_{\pm} \psi_i^*\psi_j\,|i\rangle\langle j|$.

\subsection*{Magnetic resonance analogy}

Now we are going to exploit the formal analogy between Eq.~\ref{BlochEOM} and the equations of magnetic resonance of a single spin in radio-frequency magnetic spectroscopy. This theory has been outlined by \cite{Redfield57,Bloch57,Freed68,Leporini92,Leporini95,AbragamBK,SlichterBK}. It has been lately applied to studies of decoherence problems in qubits \cite{Loss2004,Burkard2004,Paris2016}.
The approach goes as follows. From Eq.~\ref{EQTLS} we can write a von Neumann equation for the density matrix defined above, i.e. $
i\dot{\rho}(t)=\bigl[\mathrm{H}(t),\rho(t)\bigr]=\mathcal{L}(t)\rho(t)$.
Switching to the interaction picture we get rid of the deterministic part of the Hamiltonian for the time being. Thus we set $\rho'(t)= \mathrm{U}^{\dagger}(t)\rho(t)\mathrm{U}(t)\,$, $\mathrm{H}'(t)=\mathrm{H}_s'(t)$, where $\mathrm{H}_s(t)$ is the stochastic-only part of the starting Hamiltonian (see Eq.~\ref{EQStochastic}) and $\mathrm{U}(t)$ is defined by $\frac{d\mathrm{U}(t)}{dt}=-i\langle\mathrm{H}(t)\rangle\mathrm{U}(t)$ with $\mathrm{U}(0)=\openone$ \cite{AbragamBK}. The previous equation then becomes \footnote{Since $\frac{d}{dt}\mathrm{U}^{-1}=i\mathrm{U}^{-1}\langle\mathrm{H}\rangle$ (which follows from the differential equation for $\mathrm{U}(t)$) we have indeed (plugging the diff. eq. defining $\mathrm{U}$ and $\mathrm{U}^{-1}$ and the von Neumann equation) $\dot{\rho}'(t)=\dot{\mathrm{U}}^{-1}\rho\mathrm{U}+\mathrm{U}^{-1}\dot{\rho}\mathrm{U}+\mathrm{U}^{-1}\rho\dot{\mathrm{U}}=i\mathrm{U}\langle\mathrm{H}\rangle\rho\mathrm{U}^{-1}-i\mathrm{U}\mathrm{H}\rho\mathrm{U}^{-1}+i\mathrm{U}\rho\mathrm{H}\mathrm{U}^{-1}-i\mathrm{U}\rho\langle\mathrm{H}\rangle\mathrm{U}^{-1}=-i\bigl[\mathrm{H}_s',\rho'\bigr]\,$.}
\begin{equation}
i\dot{\rho}'(t)=\mathcal{L}'(t)\rho'(t)=\bigl[\mathrm{H}_s'(t),\rho'(t)\bigr]\,.
\end{equation}
Now we seek a solution by iteration \cite{VanKampenBK,Skinner87} we write
\begin{align}
\rho'(t)&=\rho(0)-i\int_0^t\mathcal{L}'(t_1)\rho(0)dt_1-\\
&-\int_0^t\int_0^{t_1}\mathcal{L}'(t_1)\mathcal{L}'(t_2)\rho'(t_2)dt_1dt_2
\end{align}
Going on like this we get to \cite{VanKampenBK,Skinner87}
\begin{equation}
\rho'(t)=\mathbb{Y}(t|0)\rho(0)\,\implies\,\langle\rho'(t)\rangle=\langle\mathbb{Y}(t|0)\rangle\rho(0),
\end{equation}
since $\rho(0)=\rho'(0)$ is not random, and where we have introduced the non-local kernel $\mathbb{Y}(t|0)=\openone+\sum_{n=1}^{+\infty}(-i)^n\int\dots\int\mathcal{L}'(t_1)\dots\mathcal{L}'(t_n)dt_1\dots dt_n$. Differentiating and assuming that $\langle\mathbb{Y}(t|0)\rangle$ is invertible then
\begin{equation}
\langle\dot{\rho}'(t)\rangle=\langle\dot{\mathbb{Y}}(t|0)\rangle\rho(0)=\langle\dot{\mathbb{Y}}(t|0)\rangle\langle\mathbb{Y}(t|0)\rangle^{-1}\langle\rho'(t)\rangle,
\label{EQY}
\end{equation}
where $\mathbb{K}'(t)\equiv\langle\dot{\mathbb{Y}}(t|0)\rangle\langle\mathbb{Y}(t|0)\rangle^{-1}$ is a non-stochastic superoperator by construction, since it connects averaged quantities. We expand $\mathbb{K}'(t)$ in orders of $G$ and 
truncate this series at the second order. Utilizing $\langle\mathcal{L}_s(t)\rangle=0$ we see that the cumulants simplify indeed to the moments of $\mathcal{L}_s(t)$,
\begin{align}
&\langle\dot{\rho}'(t)\rangle=\mathbb{K}^{\rm{II}}(t)\langle\rho'(t)\rangle\,,\label{EQCE}\\
&\mathbb{K}^{\rm{II}}(t)=-\int_0^{t}dt'\langle\mathcal{L}'_s(t)\mathcal{L}'_s(t-t')\nonumber\rangle .
\end{align}
This is equivalent to the Born approximation in open quantum systems. For that to be a good approximation it is sufficient that $|\mathcal{L}_s(t)|\tau_c\ll 1$, since this is the relative error between successive orders.
The above requirement is equivalent to demanding that the time scales of the noise and of the deterministic evolution are well separated, i.e., that $\langle\mathcal{L}(t)\rangle$ varies significantly on a time scale which is much slower than the noise memory time. Therefore, we will say that the truncated series is a coarse grained description of the full dynamics of $\rho(t)$.

Since the time scales are well separated and the noise spectrum exponentially decays we can extend the limit of integration to $+\infty$, therefore we can see that our expansion is in fact an expansion in $G\tau_c\ll 1$, since $\tau_c$ is the width of the interval where the integrand gives an important contribution. The extension of the integration interval leads to:
\begin{align}
\langle\dot{\rho}'(t)\rangle&=\mathbb{K}^{\rm{II}}(+\infty)\langle\rho'(t)\rangle\nonumber\\&=-\int_0^{+\infty}dt'\langle\mathcal{L}'_s(t)\mathcal{L}'_s(t-t')\rangle\langle\rho'(t)\rangle\,.
\end{align}
This is the Markov approximation of our non-Markovian process. We arrive at:
\begin{align}
\langle\dot{\rho}(t)\rangle&=-i\bigl[\langle\mathrm{H}(t)\rangle,\langle\rho(t)\rangle\bigr]-\int_0^{+\infty}\langle\bigl[\mathrm{H}_s(t),\label{EQMasterS}\\
&\bigl[\mathrm{U}^{\dagger}(t-t',t)\mathrm{H}_s(t-t')\mathrm{U}(t-t',t),\langle\rho(t)\rangle\bigr]\bigr]\rangle dt'\,,\nonumber
\end{align}
where $\mathrm{U}(t-t',t)\equiv\mathrm{U}(t-t')\mathrm{U}^{\dagger}(t)$. We approximate $\mathrm{U}(t-t',t)$ with $\exp{\bigl(i\mathrm{H}_0\,t'\bigr)}$, which is justified if $D\tau_c\ll1$ \cite{AbragamBK}. That means basically that in the relaxation term of the second order cumulant equation the time evolution operator is $\mathrm{U}(t)=\exp\bigl(-i\mathrm{H}_0 t\bigr)$ (to show this it is sufficient to plug in $t'=t$ in the previous relation and invert). This procedure is customary in radio frequency magnetic resonance in liquids and it is usually called the nonviscous-liquid approximation \cite{AbragamBK}. To compute the relaxation times we will make use of \cite{AbragamBK}:
\begin{widetext}
\begin{align}
\langle\dot{\rho}(t)\rangle+i\bigl[\langle\mathrm{H}(t)\rangle,\langle\rho(t)\rangle\bigr]\approx&-\int_0^{+\infty}\langle\bigl[\mathrm{H}_s(t),\bigl[e^{-i\mathrm{H}_0 t'}\mathrm{H}_s(t-t')e^{i\mathrm{H}_0 t'},\langle\rho(t)\rangle\bigr]\bigr]\rangle dt'\label{EQAbragam}\\
=&-\,e^{-i\mathrm{H}_0 t}\int_0^{+\infty}dt'\langle\bigl[\mathrm{H}^*_s(t),\bigl[\mathrm{H}^*_s(t-t'),\langle\rho^*(t)\rangle\bigr]\bigr]\rangle e^{i\mathrm{H}_0 t}\approx\mathrm{U}(t)\langle\dot{\rho}'(t)\rangle\mathrm{U}^{\dagger}(t)\,.\nonumber
\end{align}
\end{widetext}
This basically means that $\langle\dot{\rho}'(t)\rangle(t)\approx\langle\dot{\rho}^*(t)\rangle(t)$ if we remember to sum the correct first order term when we turn back to the Schrödinger picture.
Therefore, our approximation is equivalent to assuming $\mathbb{K}^{\rm{II}}(+\infty)\approx-\int_0^{+\infty}dt'\langle\mathcal{L}^*_s(t)\mathcal{L}^*_s(t-t')\rangle$ (please note that this expression still depends on $t$) in Eq.~\eqref{EQCE}, the superoperator $\mathcal{L}^*_s=\bigl[e^{i\mathrm{H}_0 t}\mathrm{H}_s(t)e^{-i\mathrm{H}_0 t},\,\circ\,\bigr]$ being
\begin{align}
\mathcal{L}^*_s(t)=&\frac{\Gamma_d(t)}{2\Omega}\Big\{\varepsilon_0\bigl[\sigma_z,\,\circ\,\bigr]-\Delta e^{-i\Omega t}\bigl[\sigma_+\,,\,\circ\,\bigr]\nonumber\\
&-\Delta e^{i\Omega t}\bigl[\sigma_-\,,\,\circ\,\bigr]\Big\}\,.\label{EQLiouville}
\end{align}
Now we split the density matrix in its spin components $\{\sigma_z,\sigma_+,\sigma_-\}$, i.e. $\boldsymbol{\mathrm{r}}=\rm{Tr}\bigl[\rho^S(t)\boldsymbol{\sigma}\bigr]=\rm{Tr}\bigl[\rho^H(0)\boldsymbol{\sigma}(t)\bigr]$.

The equation of motion for $\mathrm{r}_{\alpha}(t)
$ with $\alpha=+,-,0$, where we renamed $\mathrm{r}_z$ as $\mathrm{r}_0$, are then (averages of these components are intended, but we do not write them to ease the notation),
\begin{align}
2\Omega^2\,\dot{\mathrm{r}}'_+(t)= & -\bigl[2\varepsilon_0^2 k_0+\Delta^2k_-\bigr]\mathrm{r}'_+ \label{EQBlochInt}\\
&+ \Delta^2k_+e^{-i2\Omega t}\mathrm{r}'_- -2\Delta\varepsilon_0 k_+e^{-i\Omega t}\mathrm{r}'_0\,,\nonumber\\
2\Omega^2\,\dot{\mathrm{r}}'_-(t)= & -\bigl[2\varepsilon_0^2 k_0+\Delta^2k_+\bigr]\mathrm{r}'_-\\
& + \Delta^2k_-e^{i2\Omega t}\mathrm{r}'_+ -2\Delta\varepsilon_0 k_-e^{i\Omega t}\mathrm{r}'_0\,,\nonumber\\
2\Omega^2\,\dot{\mathrm{r}}'_0(t)= & -\Delta^2\bigl[k_-+k_+\bigr]\mathrm{r}'_0\\
& -\Delta\varepsilon_0 k_0e^{i\Omega t}\mathrm{r}'_+ -\Delta\varepsilon_0 k_0e^{-i\Omega t}\mathrm{r}'_-\,.\nonumber
\end{align}
Here we wrote $k_{\alpha}=\int_{0}^{+\infty}\langle\Gamma_d(t)\Gamma_d(t-t')\rangle e^{i\alpha\Omega \tau} dt'=G\times\frac{1+i\alpha\Omega\tau_c}{\alpha^2\Omega^2\tau_c^2+1}$ and $\mp$ is a shorthand for $\{-1\,,\,1\}$, thus $\alpha\in\{-1\,,\,0\,,\,1\}$. Furthermore, we dropped the subscript of $\varepsilon_0$. We rename these terms using Redfield's notation \cite{Redfield57}.
\begin{equation}
\langle\dot{\mathrm{r}}'_\alpha(t)\rangle=-\sum_\beta \exp\bigl[i(\beta-\alpha)\Omega t\bigr]\mathcal{R}_{\alpha,\beta}\,\langle\mathrm{r}'_\beta(t)\rangle .
\label{EQRedfield}
\end{equation}

Now Eqs.~\eqref{EQBlochInt}--\eqref{EQRedfield} can be solved perturbatively after Laplace-transforming them. In this way we finally obtain the relaxation times and the Lamb's shift $\delta\Omega$, in closed form. These are found to be:
\begin{align}
T_1^{-1}&=\mathcal{R}_{0,0}-2\,\mathrm{Re}\biggl(\frac{\mathcal{R}_{0,+}\mathcal{R}_{+,0}}{i\Omega+\mathcal{R}_{+,+}}\biggr),\label{EQTime1}\\
T_2^{-1}&=\mathrm{Re}\biggl(\mathcal{R}_{+,+}+\frac{\mathcal{R}_{0,+}\mathcal{R}_{+,0}}{i\Omega-\mathcal{R}_{0,0}}\biggr),\label{EQTime2}\\
\delta\Omega&=\mathrm{Im}\biggl(\mathcal{R}_{+,+}+\frac{\mathcal{R}_{0,+}\mathcal{R}_{+,0}}{i\Omega-\mathcal{R}_{0,0}}\biggr)-\frac{\bigl|\mathcal{R}_{+,-}\bigr|^2}{2\Omega}. \label{EQTime3}
\end{align}

\begin{figure}
\begin{center}
\includegraphics[scale=0.31]{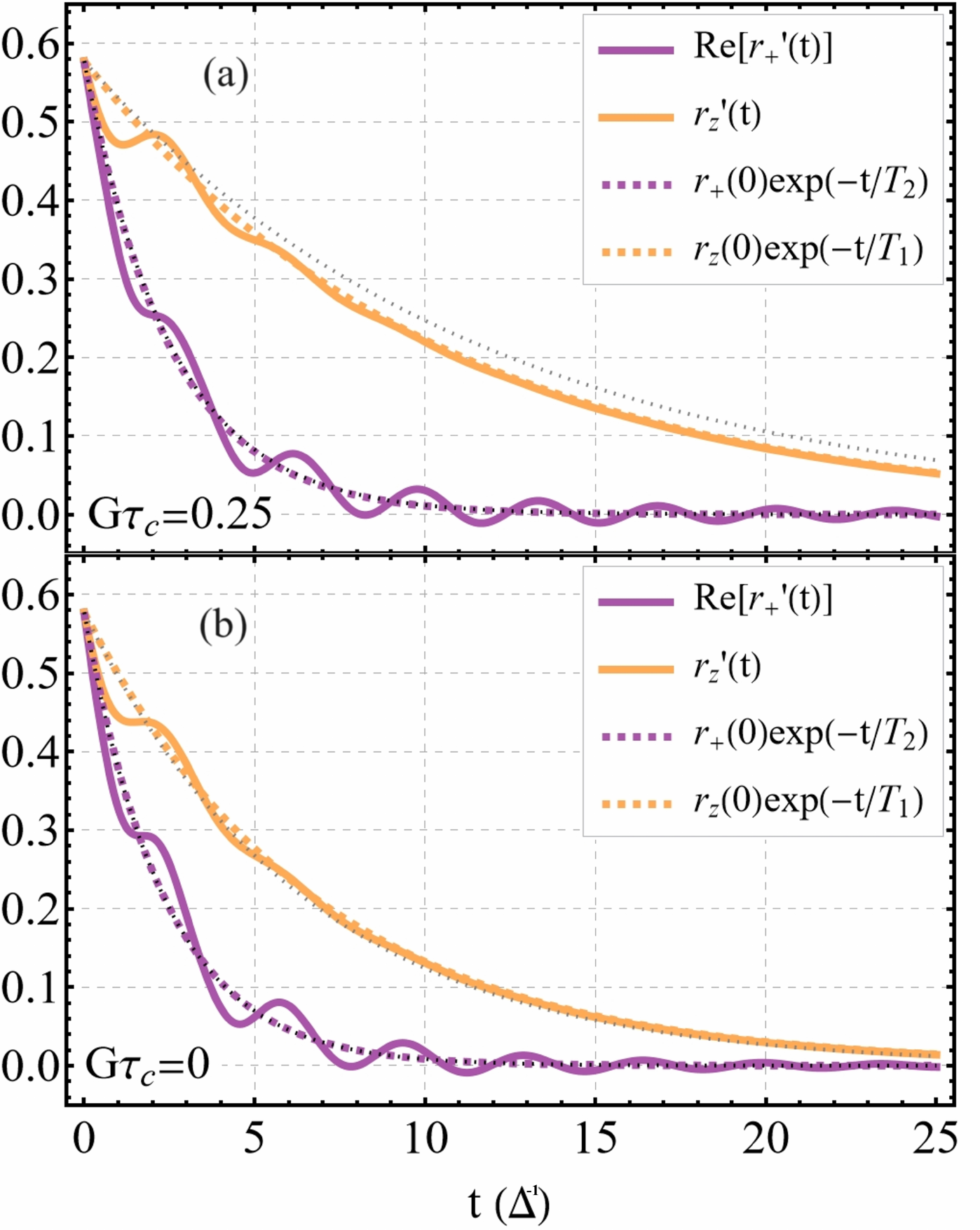}
     \caption{Decay of the polarization $\mathrm{r}'_z={\rm Tr}(\sigma_z\rho)$ of the Bloch vector (orange), and of the coherences $\rm{Re}(\mathrm{r}'_+)={\rm Tr}(\sigma_+\rho)$ (purple), for $\varepsilon_0=1.5\Delta, G=0.5\Delta$. Solid lines are the plot of the numerical solution to the Redfield equation \eqref{EQRedfield}. Dashed colored lines picture the exponential decay with times given by Eqs.~\eqref{EQTime1},\eqref{EQTime2} ($\mathrm{r}'_i(0)=\mathrm{r}_i(0)$). Dotted gray and black lines represent the corresponding plot neglecting the oscillating terms on the right hand side of Eq.~\eqref{EQBlochInt} (secular approximation). (a) Correlated noise, $G\tau_c=0.25$, and (b) white noise limit, $G\tau_c=0$.}
     \label{ExpDecay}
\end{center}
\end{figure}

These results agree with the ones reported by Refs.~\onlinecite{Leporini92,Leporini95}.
Here, the $\mathcal{R}_{i,j}$ constants are given by:
\begin{align}
\mathcal{R}_{0,0}&=\frac{G\,\Delta^2}{\Omega^2(1+\Omega^2\tau_c^2)},\\
\mathcal{R}_{0,+}&=\frac{G\,\Delta\varepsilon_0}{2\Omega^2},\\
\mathcal{R}_{+,0}&=\frac{G\,\Delta\varepsilon_0(1+i\Omega\tau_c)}{\Omega^2(1+\Omega^2\tau_c^2)},\\
\mathcal{R}_{+,+}&=\frac{G}{2}\,\biggl[\frac{2\varepsilon_0^2}{\Omega^2}+\frac{\Delta^2(1-i\Omega\tau_c)}{\Omega^2(1+\Omega^2\tau_c^2)}\biggr],\\
\mathcal{R}_{+,-}&=-\frac{G\,\Delta^2(1+i\Omega\tau_c)}{2\Omega^2(1+\Omega^2\tau_c^2)}.
\end{align}

Now substituting these results into Eqs.~\eqref{EQTime1}, \eqref{EQTime2}, and \eqref{EQTime3}, we see that correlations turn out to be fourth order corrections in those expressions for the relaxation times and the frequency shift. In Appendix \ref{sec:4thorder} we show that the fourth-order term of the cumulant expansion \eqref{EQCE} is vanishing due to the Gaussianity of the noise process and its zero avarage. In Fig.~\ref{ExpDecay} we plot $\mathrm{Re}[\mathrm{r}'_+]$ (purple) and $\mathrm{r}'_z$ (orange) in the interaction picture as a function of time. The solid lines in the plot are the numerical solution of Eq.~\eqref{EQBlochInt} while the dashed exponentials are the result of the perturbative solution given by Eqs.~\eqref{EQTime1} and \eqref{EQTime2}. The gray and black dotted lines are the solution obtained by discarding the correlations from the beginning, or by applying the secular approximation to Eq.~\eqref{EQRedfield}. We must stress also that these plots do not show the common exponential decay arising from the friction induced damping that we dropped passing from Eq. \ref{EQDrivQubit} to Eq. \ref{EQStochastic}. Theoretically, the only constraint we have on the noise parameter are the one imposed by the condition of validity of the cumulant expansion, i.e. $G\tau_c\ll 1$ \cite{VanKampenBK} and of the non-viscous liquid approximation $D\tau_c\ll 1$\cite{AbragamBK}. Obviously, these parameters need to be fine tuned in experimental realisations \ref{SecEXP}, also to account for other time-scales that can be relevant in those physical systems.

Equations~\eqref{EQTime1}, \eqref{EQTime2}, and \eqref{EQRedfield} imply that $T_2^{-1}\ne (2T_1)^{-1}$,  the difference amounting to the pure dephasing rate
\begin{equation}
T_{\phi}^{-1}=
(T_2)^{-1}-(2T_1)^{-1}
\approx\frac{G\varepsilon_0^2}{\,\Omega^2}+\frac{G^2\Delta^2\varepsilon_0^2\tau_c}{\Omega^4(1+\Omega^2\tau_c^2)}\,,
\label{EQDeph}
\end{equation}
where we neglected $\mathcal{R}_{0,0}$ and $\mathcal{R}_{+,+}$ in the denominator of Eqs.~\eqref{EQTime1} and \eqref{EQTime2}, since they lead to corrections of higher order in $G$.
This analytically approximate result and the graphs of Fig.~\ref{ExpDecay} showing exponential decay with different characteristic times demonstrate that the addition of noise to the parametric driving of this coupled classical oscillator system induce a dephasing dynamics, resolving the problem of equal relaxation times pointed out in the previous models by Refs.~\onlinecite{Novotny2014,Nori2018}. It is evident from the formula \eqref{EQDeph} above that the dephasing rate vanishes for $\varepsilon_0=0$. We will discuss briefly about the possibility for measurement of this dephasing time in section \ref{SecEXP}.

\begin{figure}
\begin{center}
\includegraphics[scale=0.4]{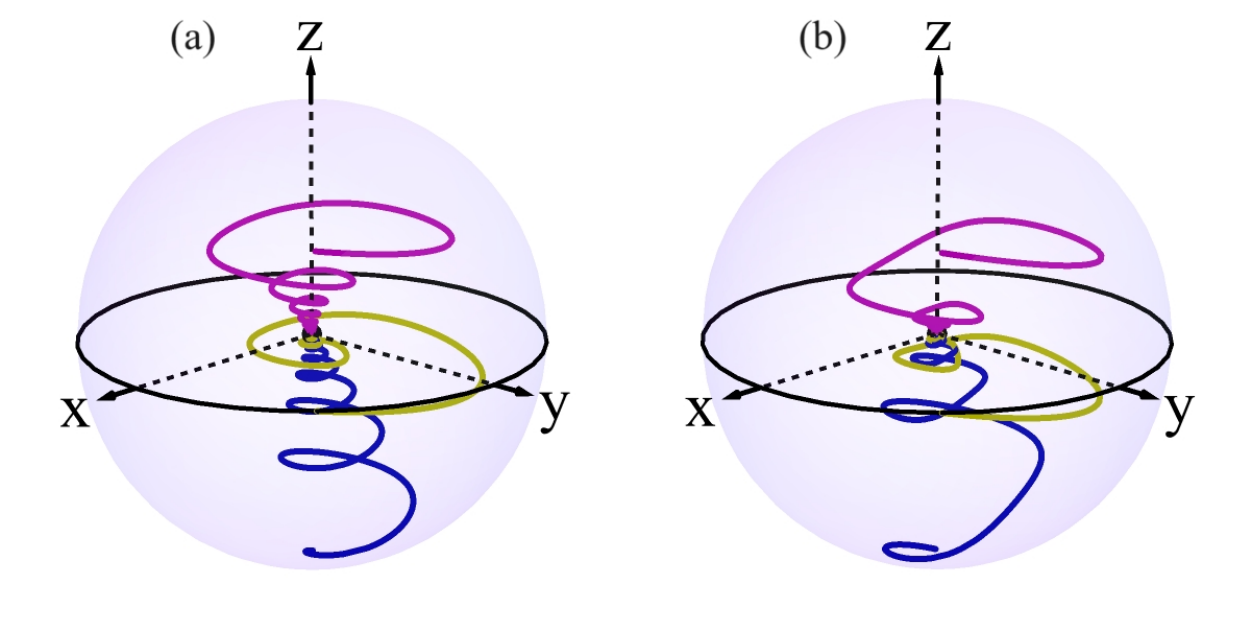}
\caption{Trajectory of the BV on the unit Bloch sphere for different initial conditions, i.e., $\boldsymbol{\mathrm{r}}(0)=\bigl[1/\sqrt{3}\,,\,1/\sqrt{3}\,,\,1/\sqrt{3}\bigr]^T$ (purple), $\boldsymbol{ \mathrm{r}}(0)=\bigl[1/\sqrt{2}\,,\,1/\sqrt{2}\,,\,0\bigr]^T$ (yellow), and $\boldsymbol{ \mathrm{r}}(0)=\bigl[0\,,\,0\,,\,-1\bigr]^T$ (blue). Both are plotted in the case of resonant driving $\omega=\Delta$ and for (a) $G=0.3\Delta$, $D=0.5\Delta$ and $\Delta\tau_c=0.5$, and (b) $G=0.3\Delta$ and $D=\Delta$, $\tau_c=0$. All BV trajectories decay to the infinite temperature state at the center of the Bloch sphere.}
\label{Sfere}
\end{center}
\end{figure}

In Fig.~\ref{Sfere} we show the BV dynamics on the unit Bloch sphere. These are plotted using Eq.~\eqref{EQBlochInt} after transforming to the lab frame (x,y,z) and to the Schr\"odinger picture. Both panels represent the time evolution of the BV for the resonantly driven TLS. The three different initial conditions give rise to three trajectories on the sphere: $\boldsymbol{\mathrm{r}}(0)=\bigl[1/\sqrt{3}\,,\,1/\sqrt{3}\,,\,1/\sqrt{3}\bigr]^T$ (purple), $\boldsymbol{\mathrm{r}}(0)=\bigl[1/\sqrt{2}\,,\,1/\sqrt{2}\,,\,0\bigr]^T$ (yellow), $\boldsymbol{\mathrm{r}}(0)=\bigl[0\,,\,0\,,\,-1\bigr]^T$ (blue). Fig. \ref{Sfere}(a) shows the system subject to noise with a memory time $G\tau_c=0.15$ and Fig. \ref{Sfere}(b) for memoryless noise, i.e. $G\tau_c=0$.

\subsection*{Why infinite temperature?}

It has already been shown by Kubo \cite{Kubo89} that the stationary states of systems governed by stochastic Hamiltonians are the infinite temperature states, i.e. $\rho\,(t\to+\infty)\propto\openone$. As it is evident from Fig.~\ref{Sfere} the BV of our system always collapses to the center of the sphere, i.e. infinite temperature state, so the previous result applies here as well. Hereafter we explain why this is the case for our system, following \cite{Franco2019}.

We recast the Redfield equation in 
a Kossakowski-Lindblad form \cite{Gorini76,Lindblad76}. Since we have seen in this section that the oscillating terms do not give an important contribution to the decay of the BV, we ignore those terms here, therefore applying a secular approximation. the resulting master equation, under the Born-Markov approximation, from Eqs.~\eqref{EQCE} and \eqref{EQLiouville} is then,
\begin{align}
\langle\dot{\rho}'(t)\rangle&=\mathbb{K}^{\rm{II}}(t)\langle\rho'(t)\rangle\approx-\frac{G^2}{4\Omega^2}\Bigl(\varepsilon^2_0k_0[\sigma_z,[\sigma_z,\langle\rho'\rangle]]\\&+\Delta^2 k_+[\sigma_+,[\sigma_-,\langle\rho'\rangle]]+\Delta^2 k_-[\sigma_-,[\sigma_+,\langle\rho'\rangle]]\Bigr).\nonumber
\end{align}
Then, we transform back to the Schr\"odinger picture and apply the secular approximation.
\begin{widetext}
\begin{equation}
\langle\dot{\rho}\rangle\approx- i\bigl[\langle\mathrm{H}\rangle,\langle\rho\rangle\bigr]-\frac{G^2}{4\Omega^2}\Bigl[2\varepsilon^2_0k_0(\langle\rho\rangle-\sigma_z\langle\rho\rangle\sigma_z)+\Delta^2\mathrm{Re}(k_+)(\langle\rho\rangle-2\sigma_+\langle\rho\rangle\sigma_--2\sigma_-\langle\rho\rangle\sigma_+)+i\Delta^2\mathrm{Im}(k_+)[\sigma_z,\langle\rho\rangle]\Bigr].
\label{EQLindblad}
\end{equation}
\end{widetext}
Here, the first term in the square brackets describes dephasing, while the second term leads to quantum jumps, and the third term is the Lamb shift.
Now since the rates for upward and downward transitions are the same, i.e. $\gamma_{\pm}=G^2\Delta^2\mathrm{Re}(k_+)/2\Omega^2$, we can ascribe an infinite effective temperature to the system, since we do not have detailed balance \cite{Redfield65,Franco2019}. Clearly this is not a true temperature since we do not describe a thermal bath, with which the system can thermalize. Nonetheless, it is a nontrivial consequence of our model and indeed an interesting limit of the discussed quantum-classical analogy.

\section{Proposals for experimental tests}
\label{SecEXP}

We now briefly analyze some experimental setups in which the discussed quantum analog could be implemented.
Among the many possible experimental test-beds we chose levitated nanoparticles and nanomechanical string resonators, since these systems are currently attracting much interest thanks to the great degree of control and isolation from the environment that they guarantee \cite{Aspelmayer2020,NovotnyNature2021,Stickler2020,NovotnyLevi2021}. Nonetheless, a realization with macroscopic setups could also be within reach \cite{Kiselev2022}, favouring the possibility of a macroscopic quantum analogy of unprecedented depth.

\subsection{Levitated nanoparticles}
In levitation experiments dieletric particles are trapped in the focus of a laser beam, using radiation pressure. The harmonic modes of oscillation  are the the center of mass coordinate of the particle, with the equilibrium position being the focus of the laser. The oscillation along the laser beam can be frozen so that the motion is effectively restricted to the two modes belonging to the focal plane. The two modes have different eigenfrequencies due to the polarization of the laser which creates an elliptically-shaped potential well. These modes can be coupled by harmonically varying the polarization direction of the trapping laser \cite{Novotny2016,Novotny2017}. Therefore, periodically varying the polarization angle according to $\theta=\delta\cos(\omega t+\varphi)$  realizes a parametric driving of the coupled oscillator system. When the frequency of this harmonic driving is in resonance with the mode splitting, the strong coupling regime 
can be reached and Rabi oscillations can be implemented, in which the energy can be exchanged coherently between the modes.

Stochastic terms in the parametric drive can be engineered utilizing, for instance, laser intensity fluctuations \cite{Quidant2016}. Nonetheless, in order to obtain a dephasing dynamics in our system it is essential to have an off-set parameter such as $\varepsilon_0$ in Eq.~\eqref{EQStochastic} that shifts the working point of the system slightly off the center of the avoided crossing. This makes the driving term appear as $\theta=\delta\cos(\omega t+\varphi)+\delta_0$, with $\delta_0\sim\delta$. Thus, one needs to be able to induce some asymmetry between the mode eigenfrequencies independently of the coupling/driving mechanism. Otherwise a shift of the polarization angle of the trapping laser in this system will only lead to a redefinition of the mode frequencies. However, in these systems the modes eigenfrequency are defined by the polarization of the laser \cite{Toros2018}. Therefore, we think that in this setup the effect  described here cannot be observed. We cannot exclude that more complicated configurations exist, in which this mode frequency asymmetry can be adiabatically tuned, independently from the driving. In this way, taking the equation (8) from Ref.~\onlinecite{Novotny2017} we can map it to an equation similar to ours by the coordinate transformation $\begin{bmatrix}a'\\b'\end{bmatrix}=\exp\bigl(i\frac{\theta}{2}\sigma_y\bigr)\begin{bmatrix}a\\b\end{bmatrix}$, where $\theta=\arctan(\delta_0/\omega_{\delta})$.

\subsection{Nanomechanical string resonators}
Considering nanoscale string resonators we propose exploiting the dielectric protocols for coupling and driving the in-plane and out-of-plane modes of a nanomechanical doubly clamped beam \cite{Unterreithmeier2009,SilvanSchmid2010,Weig2012,Weig2012APL,Weig2013,Seitner2016,Seitner2017}. On the other hand, similar physics can be implemented in nanomechanical systems with a piezoelectrical driving setup \cite{Yamaguchi2012,Okamoto2013,Yamaguchi2017}.
Considering a doubly clamped string resonator, we have that the two modes (in-plane and out-of-plane) have different eigenfrequencies, that is $k_1\ne k_2$, due to the rectangular cross-section. The oscillations of these two modes will be driven dielectrically. The parametric driving term can be written as $\Delta k\approx -\frac{\epsilon_0(\epsilon_d-1)}{2\pi}\alpha^2LF(x)\bigl[V_{dc}^2+2V_{dc}V_{ac}(t)\bigr]=C\bigl[V_{dc}^2+2V_{dc}A\cos (\omega t)\bigr]$ \cite{SilvanSchmid2010}, where $\epsilon_0$ and $\epsilon_d$ are the dielectric permittivity of vacuum and of the dielectric composing the beam (e.g. SiN), $L$ is the total length of the beam, $\alpha$ is the attenuation parameter of the electric field inside the dielectric and $F(x)$ depends on the geometry of the device. In this way the system is governed by equations of motion completely analogous to Eq.~\eqref{EQMot}, where, in the language of Ref.~\onlinecite{Weig2013}, the force $f(t)$ is the radiofrequency drive that initializes the system, $h$ is the coupling provided by the cross derivatives of the electric field generated by the gold electrodes, and $\omega_1=\sqrt{2k_1/m}$ and $\omega_2=\sqrt{2k_2/m}$ are the frequencies of the in-plane and out-of-plane modes, where $m$ is the total mass of the beam. In our language then $\varepsilon_0\approx C(V_{dc}-V_0)^2$ is proportional to the dc part of the voltage driving, where $V_0$ is the voltage corresponding to the centre of the level splitting, and $\varepsilon'(t)$ is given by the ac part $CA(V_{dc}-V_0)V_{ac}\cos(\omega t)$.
In this way the coupled modes of these nanomechanical string resonators are described by the same equations that we used and we think they can simulate the dephasing dynamics here described.

\section{Conclusions}

We showed how the analogy between classical coupled oscillators and quantum TLS can be extended to include a decoherent dynamics of the TLS. We computed the relaxation times for BV components and demonstrated that a pure dephasing $T_{\phi}$ (Eq.~\eqref{EQDeph}) time appears in the TLS dynamics, as a consequence of the addition of noise to the parametric driving of the classical system. This analogy can be exploited to effectively simulate or probe quantum decoherence dynamics and dissipation via classical means, resulting in an effective simulation of decoherence. However, the simulation in its present form cannot grasp decay processes such as spontaneous emission. Another drawback of the model here presented is that it scales exponentially with respect to the size of the corresponding quantum image. In fact, to simulate a collection of $N$ qubits, $2^N$ coupled classical oscillators are required \cite{Briggs2013}. This poses obvious limitations on the practical utility of this approach.

Simulation of open-quantum-system dynamics by addition of classical noise to analog, more controllable, physical system has spurred much interest lately \cite{DelCampo2017,Franco2019}. In particular, Ref.~\onlinecite{Franco2019} studied how classical noise can mimic quantum dissipation and derived very clear analogies between the two frameworks, considering for instance the spin-boson problem. In a more complicated framework Ref.~\onlinecite{DelCampo2017} studied how a wide class of master equations for many-body systems can be simulated in easily controllable systems subject to classical noise.
Our analysis constitutes a simplification of these earlier attempts, since it does not require any quantum system at all, therefore enabling a purely classical simulation of quantum dissipation problems, that in principle can be implemented in systems even beyond the nanoscale. Albeit some other dissipative phenomena, e.g. spontaneous emission, are still out of reach for these kinds of simulation due to fundamental issues \cite{Burgarth2017,Franco2019}. Also in the system  discussed here, the equality between the upward and downward transition rate in Eq.~\eqref{EQLindblad} withhold the possibility to simulate spontaneous decay processes.

Notably, the system investigated here can be valuable also for frequency noise detection in the aforementioned mechanical systems. Usually, this kind of noise is hard to investigate because of the interplay of the thermal fluctuations. However, we see that in the system treated here these fluctuations are fully decoupled from the frequency noise and do not affect the relaxation times that we found (see Appendix \ref{sec:non-hom}). Therefore, one could measure frequency noise features from the expected decay times.

\section*{Acknowledgements}
We gratefully acknowledge funding from the Deutsche Forschungsgemeinschaft Project No.~425217212, SFB 1432.

\appendix

\section{Non-homogeneus forcing case}
\label{sec:non-hom}
We now analyze what happens considering the full case, i.e. the one with inhomogeneous forcing. We will be assuming that in this case Eq. \ref{EQTLS} can be easily modified by adding a vector with two noisy complex components, i.e. $\boldsymbol{f}(t)=[\,f_1(t)\,f_2(t)\,]^T$. This is not too far from the treatment of Ref. \onlinecite{Novotny2014}, even if we should also account for the non-stationarity of the time factors introduced by the SVEA, that we will nonetheless neglect for simplicity. Therefore we will consider $f_{1,2}(t)$ as a noise realisation of a stationary Gaussian process. Let us consider then the larger vectorial (let us call it a combined Hilbert-Liovillean) space, where the following vector lives:
\begin{equation}
\boldsymbol{\Psi}=
\begin{bmatrix}
\psi_1\\
\psi_2\\
\psi_1^*\psi_1\\
\psi_1^*\psi_2\\
\psi_2^*\psi_1\\
\psi_2^*\psi_2
\end{bmatrix}\nonumber
\end{equation}
So that we can treat everything on the same foot. Assuming that both $D$ and $G$ are small parameters with respect to $\Omega=\sqrt{\Delta^2+\epsilon^2_0}\,$, the dynamics of this large vector can be described in the following way \cite{Roerdink81}:
\begin{align}
\langle\boldsymbol{\dot{\Psi}}(t)\rangle=&\,\mathcal{K}(t)\langle\boldsymbol{\Psi}(t)\rangle\label{EQ10}\\
=&\begin{bmatrix}
\mathrm{K}(t) & \varnothing_{2\times 4} \\
\mathbb{F}_{4\times 2}(t) & \mathbb{K}(t)
\end{bmatrix}\langle\boldsymbol{\Psi}(t)\rangle \nonumber\\
&+
\begin{bmatrix}
\langle\boldsymbol{f}(t)\rangle\\
\int dt'\langle\tilde{\boldsymbol{f}}(t)e^{it'\mathrm{H}_0}\boldsymbol{f}(t-t')\rangle
\end{bmatrix}\,,\nonumber
\end{align}
where the minor $\mathbb{K}(t)$ is an operator that governs the time evolution of the second moments in the Liouvillean space. This is the analog of Eq.~\ref{EQCE}. Considering the non-homogeneous term of the equation above \ref{EQ10} (that is the six-tuple at the end of RHS), we readily see that the first two entries vanish if this external noise has zero avarage. The remaining four entries (or rows) constitute a non-homogeneous driving term that can pump the populations and the coherences, depending on the cross and self correlations of $f_1$ and $f_2$. Nonetheless, this term cannot lead to an exponential decay of the coherences.

Let us now focus on the lower left block. This is a $4\times 2$ matrix with the following form:
\begin{gather*}
\mathbb{F}(t)=\langle\tilde{\boldsymbol{f}}(t)\rangle=\langle\boldsymbol{f}^*(t)\otimes\openone_2+\openone_2\otimes\boldsymbol{f}(t)\rangle=\\
\begin{bmatrix}
\langle f_1^*(t)\rangle\openone_2\\
\langle f_2^*(t)\rangle\openone_2
\end{bmatrix}+
\begin{bmatrix}
\boldsymbol{0} && \langle\boldsymbol{f}(t)\rangle\\
\langle\boldsymbol{f}(t)\rangle && \boldsymbol{0}
\end{bmatrix}=
\begin{bmatrix}
\langle f_1^*\rangle && \langle f_1\rangle\\
0 && \langle f_1^*+f_2 \rangle\\
\langle f_2^*+f_1 \rangle && 0\\
\langle f_2\rangle && \langle f_2^*\rangle
\end{bmatrix}
\end{gather*}
Note that if $\langle\boldsymbol{f}\rangle=0$, i.e. if the fluctuation has zero average, that is the case considered in the main text, then the super-superoperator is
\begin{equation}
\mathcal{K}(t)=
\begin{bmatrix}
\mathrm{K}(t) & \varnothing_{2\times 4} \\
\varnothing_{4\times 2} & \mathbb{K}(t)
\end{bmatrix}
\end{equation}
and notably the dynamics of vector $\boldsymbol{\Psi}$ factorizes Eq.~\eqref{EQ10} into the dynamics restricted to the two subspaces. In other words, if $\boldsymbol{\Psi}\in V$ then  $V'_1\equiv\Span\{\boldsymbol{e}_1,\boldsymbol{e}_2\}$ and $V'_2\equiv\Span\{\boldsymbol{e}_3,\boldsymbol{e}_4,\boldsymbol{e}_5,\boldsymbol{e}_6\}$ are subspaces of $V=V'_1\oplus V'_2$ preserved by the dynamics. It is useful then to decompose the space $V$ as a direct product $V=V_1\otimes V_2$, where $V_1\equiv\Span\{\boldsymbol{e}_1,\boldsymbol{e}_2\}$ and $V_2\equiv\Span\{\boldsymbol{e}_1,\boldsymbol{e}_2,\boldsymbol{e}_3,\boldsymbol{e}_4\}$. The space $V_2$ is the tetradic space. Now the equation restricted to subspace $V_1$ is just the equation for the mean values of the stochastic variables, i.e. our state vector, which we will write making use of the cumulant expansion, 
\begin{equation}
\langle\boldsymbol{\dot{\psi}}\rangle=\mathrm{K}(t)\langle\boldsymbol{\psi}\rangle,
\end{equation}
where
\begin{align}
\mathrm{K}(t)=&-i\bigr{\langle}\mathrm{H}(t)\bigr{\rangle}\\-& \int_0^{+\infty}\bigr{\langle}\bigr{\langle}\mathrm{H}_1(t)\,e^{it'\mathrm{H}_0}\mathrm{H}_1(t-t')\bigr{\rangle}\bigr{\rangle}\,e^{-it'\mathrm{H}_0}dt' , \nonumber
\end{align}
and where $\langle\langle\circ\rangle\rangle$ is the cumulant. Here the cumulant sign is actually important because the perturbation is $\mathrm{H}_1(t)=\frac{D\cos{(\omega t)}+\Gamma_d(t)}{2\Omega}\bigl(\varepsilon_0\sigma_z-\Delta\sigma_x\bigr)$ with $\langle\mathrm{H}_1(t)\rangle\nequal 0$.

The dynamics in $V_2$ space is described by the following dynamic operator
\begin{align}
\mathbb{K}(t)=&-i\bigr{\langle}\mathcal{L}(t)\bigr{\rangle}\\-& \int_0^{+\infty}\bigr{\langle}\bigr{\langle}\mathcal{L}_1(t)\,e^{it'\mathcal{L}_0}\mathcal{L}_1(t-t')\bigr{\rangle}\bigr{\rangle}\,e^{-it'\mathcal{L}_0}dt'\nonumber
\end{align}
where $\mathcal{L}_1(t)=\mathcal{L}_d(t)+\mathcal{L}_s(t)$. This  is the analog of Eq.~\eqref{EQCE} \cite{VanKampenBK}. We can show that in the perturbative regime, this dynamics actually reduces to the previous result \eqref{EQAbragam} by considering that the deterministic part of the driving cancels in the cumulants. To see this, let us expand the cumulant as 
\begin{align}
& \bigr{\langle}\bigr{\langle}\mathcal{L}_1(t)\,e^{it'\mathcal{L}_0}\mathcal{L}_1(t-t')\bigr{\rangle}\bigr{\rangle}=\\
&\bigr{\langle}\mathcal{L}_1(t)\,e^{it'\mathcal{L}_0}\mathcal{L}_1(t-t')\bigr{\rangle}-\bigr{\langle}\mathcal{L}_1(t)\bigr{\rangle}\,\bigr{\langle}e^{it'\mathcal{L}_0}\mathcal{L}_1(t-t')\bigr{\rangle}\nonumber
\end{align}
that follows from the definition of the cumulant and the fact that $e^{it'\mathcal{L}_0}$ is non-stochastic.
Now we write $\mathcal{L}_1(t)=\mathcal{L}_d(t)+\mathcal{L}_s(t)$, where $\langle\mathcal{L}_s(t)\rangle=0$, which yields

\begin{equation}
\bigl{\langle}\bigl{\langle}\mathcal{L}_1(t)\,e^{it'\mathcal{L}_0}\mathcal{L}_1(t-t')\bigr{\rangle}\bigr{\rangle}=\bigl{\langle}\mathcal{L}_s(t)\,e^{it'\mathcal{L}_0}\mathcal{L}_s(t-t')\bigr{\rangle}\,,
\end{equation}
since the terms containing the deterministic driving part in the RHS cancel with each other.
With this we have for the second order cumulant expansion,
\begin{widetext}
\begin{equation}
\mathbb{K}(t)=-i\bigr{\langle}\mathcal{L}(t)\bigr{\rangle}-\frac{1}{4\Omega^2}
\begin{bmatrix}
\Delta^2(k_++k_-) && -2\Delta\varepsilon_0 k_0 && -2\Delta\varepsilon_0 k_0 && -\Delta^2(k_++k_-) \\
-2\Delta\varepsilon_0 k_+ && 2\Delta^2k_-+4\varepsilon_0^2k_0 && 2\Delta^2k_+ && 2\Delta\varepsilon_0 k_+ \\
-2\Delta\varepsilon_0 k_- && -2\Delta\varepsilon_0 k_- && 2\Delta^2k_-+4\varepsilon_0^2k_0 && 2\Delta\varepsilon_0 k_- \\
-\Delta^2(k_++k_-) && 2\Delta\varepsilon_0 k_0 && 2\Delta\varepsilon_0 k_0 && \Delta^2(k_++k_-)
\end{bmatrix},
\end{equation}
\end{widetext}
with no dependence on the driving in the dissipator.

Note that in this way we cannot be assured that the higher orders vanish because the perturbation has non-zero stochastic average. However, if we take their expressions from \cite{Fox76}, we can see by inspection that third order and fourth order terms vanish nonetheless.

\section{Fourth order calculation}
\label{sec:4thorder}
To obtain the correction due to the correlations we should then complement our analysis by a fourth order expansion of Eq.~\ref{EQY}. Thus refining our coarse-grain description by another order. Since the noise is Gaussian and stationary, odd orders in the perturbation $\mathrm{H}_s$ vanish. Using again the non-viscous liquid approximation we get, 
\begin{widetext}
\begin{align}
\mathbb{K}^{\rm{IV}} (+\infty)\approx\iiint_0^{+\infty} &dt_1dt_2dt_3\bigl[\langle\mathcal{L}^*_s(t)\mathcal{L}^*_s(t_1)\mathcal{L}^*_s(t_2)\mathcal{L}^*_s(t_3)\rangle-\langle\mathcal{L}^*_s(t)\mathcal{L}^*_s(t_1)\rangle\langle\mathcal{L}^*_s(t_2)\mathcal{L}^*_s(t_3)\rangle\nonumber\\
-&\langle\mathcal{L}^*_s(t)\mathcal{L}^*_s(t_2)\rangle\langle\mathcal{L}^*_s(t_1)\mathcal{L}^*_s(t_3)\rangle-\langle\mathcal{L}^*_s(t)\mathcal{L}^*_s(t_3)\rangle\langle\mathcal{L}^*_s(t_1)\mathcal{L}^*_s(t_2)\rangle\bigr]\, .
\end{align}
\end{widetext}
However, the time dependence of $\mathcal{L}_s(t)=\frac{\Gamma(t)}{\Omega}\bigl[\epsilon_0\sigma_z-\Delta\sigma_x,\circ\bigr]$ is restricted to the noise realization that is the only part relevant for the integral. Nonetheless, since the process is Gaussian,  $\langle\Gamma(t)\Gamma(t_1)\Gamma(t_2)\Gamma(t_3)\rangle=\langle\Gamma(t)\Gamma(t_1)\rangle\langle\Gamma(t_2)\Gamma(t_3)\rangle+\langle\Gamma(t)\Gamma(t_2)\rangle\langle\Gamma(t_1)\Gamma(t_3)\rangle+\langle\Gamma(t)\Gamma(t_3)\rangle\langle\Gamma(t_1)\Gamma(t_2)\rangle$ and the operator part being all commuting we find that the fourth order contribution vanishes altogether.

\nocite{*}
\bibliography{references}

\end{document}